# Beyond openness: Inclusiveness and usability of Chinese scholarly data in OpenAlex

Lin Zhang[1,2,3*], Zhe Cao[1,2], Jianhua Liu[4*], Nees Jan van Eck[5]

1. Center for Science, Technology & Education Assessment (CSTEA), School of Information Management, Wuhan University, Wuhan, China
2. Center for Studies of Information Resources, School of Information Management, Wuhan University, Wuhan, China
3. Centre for R&D Monitoring (ECOOM), Department of MSI, KU Leuven, Leuven, Belgium
4. Beijing Wanfang Data Ltd., Beijing, China
5. Centre for Science and Technology Studies (CWTS), Leiden University, Leiden, The Netherlands

**\*Correspondence:** Lin Zhang, Center for Science, Technology & Education Assessment (CSTEA), School of Information Management, Wuhan University, Wuhan, Hubei Province, China, Email: linzhang1117@whu.edu.cn; Jianhua Liu, Beijing Wanfang Data Ltd., Beijing, China, Email: liujh@wanfangdata.com.cn.

**ORCID:**
Lin Zhang: 0000-0003-0526-9677
Zhe Cao: 0000-0002-9379-2013
Jianhua Liu: 0000-0002-4003-8834
Nees Jan van Eck: 0000-0001-8448-4521

**Abstract**
OpenAlex, launched in 2022 as a fully open scholarly data source, promises greater inclusiveness compared to traditional proprietary databases. This study evaluates whether OpenAlex delivers on that promise by examining its coverage and metadata quality for Chinese-language journals and their articles. Using the 2023 edition of *A Guide to the Core Journals of China* (GCJC) and Wanfang Data as a benchmark, we analyze three aspects: (1) journal-level coverage, (2) article-level coverage, and (3) completeness and accuracy of metadata fields. Results show that OpenAlex indexes only 37% of GCJC journals and 24% of their articles, with substantial disciplinary and temporal variation. Metadata quality is uneven: while basic fields such as title and publication year are complete, bibliographic details, author affiliations, and cited references are frequently missing or inaccurate. DOI coverage is limited, and language information is often incorrect, with most Chinese-language articles labeled as English. These findings highlight significant challenges for achieving full inclusiveness and usability in research evaluation and related activities. We conclude with recommendations for improving data aggregation strategies, DOI registration practices, and metadata standardization to enhance the integration of local scholarly outputs into global open infrastructures.

# 1. Introduction

The *Barcelona Declaration on Open Research Information* (2024) emphasizes that research information should not only be open but also diverse and inclusive (Babini et al., 2024). In recent years, emerging bibliographic data sources such as OpenAlex, a fully open data source launched in 2022 that is freely accessible and allows unrestricted use of its data (Priem et al., 2022), have created new possibilities for the open sharing and use of research information. These open data sources have gained growing attention and support from research institutions and are increasingly explored in research and evaluation practices (Cao et al., 2025), especially as alternatives to proprietary data sources that have traditionally dominated the field. For example, Sorbonne University, in alignment with its commitment to openness in research, opted not to renew its subscription to Web of Science and Clarivate's bibliometric tools, and redirected its focus towards exploring open alternatives, with OpenAlex emerging as a prominent candidate (Sorbonne University, 2023). Similarly, the Center for Science and Technology Studies (CWTS) at Leiden University has used OpenAlex as the basis for its novel CWTS Leiden Ranking Open Edition, the first fully transparent and verifiable ranking of major universities worldwide, made possible by the unrestricted accessibility and reusability of OpenAlex data (Van Eck et al., 2024). While OpenAlex offers an unrestricted alternative to traditional proprietary data sources, the question remains: has it fully delivered on its promise of diversity and inclusiveness in research information?

In addition to being freely accessible and unrestricted in use, OpenAlex is also recognized for its broader coverage compared to many proprietary data sources. It includes information on hundreds of millions of papers, journals, authors, and institutions, as well as billions of connections between them. Researchers have empirically studied its coverage from various perspectives, such as the coverage of open access journals (Maddi et al., 2025) and the indexing of retracted papers (Ortega et al., 2024). At the same time, challenges have been observed due to the aggregation of data from sources with varying metadata quality. Delgado-Quirós et al. (2024) have shown that although OpenAlex provides a large number of publication fields (e.g., persistent identifiers, bibliographic information, type information), the completeness of some fields is limited. For example, Zhang et al. (2024) have pointed out the frequent absence of institutional information for journal articles. In addition to completeness, accuracy is another data quality issue that is of interest. For example, Céspedes et al. (2025) have found that while OpenAlex offers a more balanced linguistic coverage than Web of Science, its language metadata is not always accurate, leading to an overestimation of English-language research outputs and an underestimation of works in other languages.

Despite the growing interest in and usage of OpenAlex, relatively little attention has been paid to whether it truly fosters diversity and inclusiveness. Traditional proprietary data sources have long faced criticism for limited and biased coverage, particularly regarding publications from diverse geographical regions (Asubiaro et al., 2023), issues that open data sources like OpenAlex could help address. Limited and biased coverage affect the reliability and use of data sources for research agenda setting, scientific evaluation, international comparisons, and other purposes. Existing studies have already demonstrated OpenAlex's notable advantage over Web of Science and Scopus in terms of its coverage of African publications (Alonso-Alvarez et al., 2025) and its indexing of open access journals from underrepresented regions or lower-income countries (Maddi et al., 2025). However, systematic studies of OpenAlex's inclusiveness and usability focusing on Asian countries, a region with substantial non-English academic output (Veugelers, 2012), remain scarce. Evidence suggests that OpenAlex exhibits substantial discontinuities and systematic biases in its coverage of publications from non-English-speaking countries such as China (Zheng et al., 2025), but this finding is primarily based on an analysis of Chinese-affiliated papers indexed in OpenAlex and lacks a comparative benchmark drawn from local data sources.

China offers a particularly relevant case. Publication and citation volumes of Chinese scholars have reached the global forefront (Brainard et al., 2022; Xie et al., 2014), making China's international research output an indispensable part of global science. However, it is noteworthy that a larger proportion of China's research output is published domestically, which has long been overlooked by traditional subscription-based data sources. According to statistics from the Service Center for Societies of CAST (2024), in 2023, approximately 997,300 citable articles were published in Chinese-language scientific journals in the natural sciences, while the number of articles authored by Chinese researchers in journals indexed by the Science Citation Index was only around 728,700. Examining the coverage of China's local research output is therefore an important test case for evaluating the inclusiveness and usability of OpenAlex data from a geographical perspective.

In light of the above context, this study focuses on OpenAlex's indexing of journals published in China and in the Chinese language (hereinafter referred to as "Chinese journals"), as well as their articles, with a particular emphasis on three questions: (1) At the journal level, to what extent does OpenAlex index Chinese journals? (2) At the article level, what is the coverage of Chinese journal articles in OpenAlex? (3) How complete and accurate is the information for individual Chinese journal articles included in OpenAlex? Building on quantitative analysis, we integrate local knowledge from China to interpret the empirical findings and provide targeted recommendations for the future development of open research information sources such as OpenAlex.

## 2. Selection of Chinese journals and data sources

**2.1 Academic journals of China**

Journal evaluation practices in China can be traced back to the 1960s and have since evolved significantly, resulting in the establishment of seven prominent journal lists used for research and evaluation purposes (Huang et al., 2021). These are the *Chinese Science Citation Database* (CSCD) and the *Journal Partition Table* (JPT) from the National Science Library, Chinese Academy of Sciences; the *AMI Journal List* from the Chinese Academy of Social Sciences Evaluation Studies; the *Chinese S&T Journal Citation Report* (CJCR) from the Institute of Scientific and Technical Information of China; *A Guide to the Core Journals of China* (GCJC) from Peking University Library; the *Chinese Social Sciences Citation Index* (CSSCI) from the Institute for Chinese Social Science Research and Assessment of Nanjing University; and the *World Academic Journal Clout Index* (WAJCI) from the China National Knowledge Infrastructure.

Among the seven journal lists, the GCJC list[1] is one of the earliest, established in 1992. It covers a broad range of disciplines and is widely recognized within China. The GCJC list is the result of a research project led by Peking University Library, with participation from a dozen other university libraries and experts from related institutions. Candidate journals for this list are those published in mainland China in the Chinese language. Since 2008, the GCJC list has been updated every three years to reflect changes in the Chinese journal landscape. It serves as an important reference for publication and research evaluation in China. Accordingly, this study focuses on the 1,987 academic journals indexed in the 2023 edition of the GCJC list and their articles published across all years.

**2.2 Data sources for comparison**

This study examines the coverage of Chinese journals from the GCJC list and their articles in OpenAlex, as well as the completeness and accuracy of information provided for individual articles. While OpenAlex is the primary data source under investigation, Wanfang Data is selected as the representative domestic data source for evaluating the quality of OpenAlex's data on Chinese research outputs.

Established in 1993 and backed by the Institute of Scientific and Technical Information of China (ISTIC) under the Chinese Ministry of Science and Technology, Wanfang Data was the very first corporation in mainland China with databases of Chinese information resources as its core business. Wanfang Data, Tsinghua Tongfang Optical Disc Company (TTOD), and Chongqing VIP Information Consulting Company (VIP) are considered as the three major Chinese journal aggregators (Atwill, 2005). Although none of these aggregators achieves full coverage of the 1,987 journals selected for this study, Wanfang Data includes most of the relevant data needed for our analysis. It maintains a comprehensive database underlying its information services by integrating data from ISTIC's original document delivery system and from the National Center for Philosophy and Social Science Documentation led by the Chinese Academy of Social Sciences[2]. This integration ensures that Wanfang Data adequately includes relevant data for the journals and their publications selected for our analysis. Moreover, one of the co-authors of this paper, Jianhua Liu, has access to the complete underlying database of Wanfang Data due to her institutional position and resource privileges, which facilitates effective data collection for this research. Therefore, Wanfang Data is selected as the benchmark for comparison with OpenAlex in this study.

## 3. Coverage of Chinese journals in OpenAlex

### 3.1 Journals matching and coverage

To answer the first research question, this section examines the coverage of the 1,987 journals from the GCJC list in OpenAlex. We first obtained basic information on these journals from Wanfang Data and then matched this journal data to OpenAlex using the p-ISSN (print ISSN)[3]. ISSN (International Standard Serial Number) is a global and unique identifier for serial publications, which helps to overcome issues such as distinguishing journals with identical names and identifying spelling or language variants of the same title, thereby enabling precise identification of journals. The use of p-ISSN is appropriate here because journals listed in GCJC must have a print edition, and the main reporting channels in China typically provide only the p-ISSN of the journals.

Among the 1,987 journals included in this study, only six lacked a p-ISSN in the Wanfang data. For these journals, we manually checked and confirmed that they are not included in OpenAlex. For journals that could not be matched using p-ISSN, we conducted additional searches using both the Chinese and English journal titles provided by Wanfang, but were unable to identify more matches. This was mainly due to OpenAlex recording Chinese journal titles in irregular pinyin forms. We manually attempted to search for 20 journals using various forms of their titles but failed to match the corresponding journals in OpenAlex. Therefore, we consider the p-ISSN-based matching approach to be appropriate for this study.

Using this approach, we found that 734 (37%) of the 1,987 journals are covered in OpenAlex.

### 3.2 Disciplinary differences in journal coverage

The coverage of Chinese journals in OpenAlex, relative to Wanfang, shows notable variation across academic disciplines. Table 1 presents the number of journals per discipline, the number included in OpenAlex, and the corresponding journal coverage. As can be observed, Chinese journals in the life and natural sciences exhibit higher levels of inclusion in OpenAlex than those in the humanities and social sciences. The three disciplines with the highest journal coverage in OpenAlex are *Biological Sciences* (70%), *Medicine and Health* (64%), and *Agricultural Science* (59%).

Table 1 Coverage of Chinese journals in OpenAlex by discipline

| Disciplines | Number of journals in Wanfang | Number of journals in OpenAlex | Journal coverage in OpenAlex |
| --- | --- | --- | --- |
| Biological Sciences | 44 | 31 | 70% |
| Medicine and Health | 253 | 162 | 64% |
| Agricultural Science | 143 | 85 | 59% |
| Astronomy and Earth Sciences | 101 | 47 | 47% |
| Environmental Science and Safety Science | 29 | 12 | 41% |
| General Natural Sciences | 188 | 76 | 40% |
| Industrial Technology | 446 | 152 | 34% |
| Language and Writing | 34 | 11 | 32% |
| Economics | 156 | 47 | 30% |
| History and Geography | 37 | 9 | 24% |
| Culture, Science, Education, and Sports | 158 | 38 | 24% |
| Aerospace and Aviation | 25 | 6 | 24% |
| Transportation | 30 | 7 | 23% |
| Marxism, Leninism, Mao Zedong Thought, Deng Xiaoping Theory | 5 | 1 | 20% |
| Philosophy and Religion | 37 | 7 | 19% |
| Arts | 32 | 5 | 16% |
| Politics and Law | 66 | 10 | 15% |
| General Social Sciences | 166 | 25 | 15% |
| Literature | 37 | 3 | 8% |
| Total | 1,987 | 734 | 37% |

Note: This study uses the Chinese Library Classification to distinguish academic disciplines, as it is recognized for its authority, standardization, and applicability in the context of Chinese academic journals.

## 4. Coverage of Chinese journal articles in OpenAlex

**4.1 Disciplinary differences in article coverage**

To answer the second research question, this section examines for each journal included in this study the number of articles available in OpenAlex relative to the total number of articles published in the journal. The analysis is conducted on a per-journal basis rather than treating the entire set of articles as a single unit, primarily because data quality is closely tied to journal-specific factors such as subject area, operational mechanisms, and publication norms. Evaluating article coverage at the journal level allows for more detailed and accurate insights.

For the 734 journals covered in both OpenAlex and Wanfang Data, we used the p-ISSN to retrieve the complete set of articles from each source. According to Wanfang Data, these journals collectively published 6,453,244 articles, which we use as the benchmark for evaluating OpenAlex's coverage and data quality. OpenAlex contains 1,549,503 works for these journals. After excluding works with missing titles and non-academic content (e.g., prefaces, tables of contents, retraction notices, editorials), a total of 1,545,929 articles (99.8%) were retained for analysis.

We calculated the ratio of the number of articles indexed in OpenAlex to the number indexed in Wanfang Data for each journal across all available publication years and found that 91% of the 734 journals have an article coverage rate below 50%. This suggests that OpenAlex's coverage of Chinese journal literature is limited not only in terms of the number of Chinese domestic journals included but also in the extent to which articles from those journals are covered.

Similar to journal coverage, the coverage of articles also varies significantly across disciplines. As shown in Table 2, journals in the life and natural sciences tend to have a higher proportion of their articles included in OpenAlex compared to those in the humanities and social sciences.

Table 2 Coverage of Chinese journal articles in OpenAlex by discipline

| Disciplines | Number of articles in Wanfang | Number of articles in OpenAlex | Article coverage in OpenAlex |
|---|---|---|---|
| Aerospace and Aviation | 29,321 | 9,857 | 36% |
| General Natural Sciences | 609,156 | 249,286 | 35% |
| Philosophy and Religion | 27,941 | 9,906 | 34% |
| Biological Sciences | 218,508 | 72,604 | 31% |
| Astronomy and Earth Sciences | 274,975 | 76,172 | 29% |
| Agricultural Science | 735,289 | 169,364 | 28% |
| Industrial Technology | 1,458,769 | 399,563 | 27% |
| Transportation | 41,943 | 9,639 | 25% |
| Environmental Science and Safety Science | 99,771 | 20,830 | 24% |
| Economics | 266,955 | 50,457 | 21% |
| Medicine and Health | 2,097,189 | 390,121 | 19% |
| Culture, Science, Education, and Sports | 242,721 | 43,689 | 19% |
| Politics and Law | 47,767 | 9,301 | 18% |
| General Social Sciences | 144,431 | 19,968 | 16% |
| Marxism, Leninism, Mao Zedong Thought, Deng Xiaoping Theory | 6,118 | 918 | 15% |
| Language and Writing | 41,453 | 4,687 | 11% |
| Literature | 13,694 | 1,206 | 10% |
| Arts | 38,132 | 3,977 | 9% |
| History and Geography | 59,111 | 4,384 | 9% |
| Total | 6,453,244 | 1,545,929 | 24% |

Note: This study uses the Chinese Library Classification to distinguish academic disciplines, as it is recognized for its authority, standardization, and relevance in the context of Chinese academic journals.

To investigate whether Chinese journal articles are truly missing from OpenAlex, we examined the possibility that some articles are actually indexed but cannot be retrieved through their p-ISSN. For this analysis, we considered all articles published in the 1,987 journals included in this study that have a DOI recorded in Wanfang Data. This amounts to 7,182,093 unique DOIs. Among these, 192,108 articles could not be retrieved in OpenAlex via p-ISSN but can be matched through their DOI. Of these, 13% have no source information in OpenAlex, 79% are linked to a source but lack an ISSN, and 8% are linked to a source where the ISSN seems to be incorrect. These findings confirm that our strategy of retrieving articles via p-ISSN is generally acceptable. However, they also reveal an important weakness of OpenAlex: a significant number of articles lack source information or contain inaccurate source metadata, which limits the reliability of journal-level analyses.

**4.2 Temporal trends in article coverage**

To further explore article-level coverage, we examined the temporal distribution of OpenAlex's indexing of Chinese journal articles. Figure 1 presents the ratio of the number of articles covered by OpenAlex to those covered by Wanfang Data for the 734 Chinese journals between 1980 and 2024.

The results show that the ratios are relatively higher between 2000 and 2015. In contrast, coverage is notably lower for earlier years and more recent years. This pattern suggests that OpenAlex's indexing of Chinese journal literature is uneven over time, with a peak around 2010. This corresponds with findings reported by Zheng et al. (2025).

We also examined whether the covered articles contain a *mag_id* in OpenAlex, which indicates that the article was already included in the Microsoft Academic Graph (MAG), the predecessor of OpenAlex. This is important because it shows whether OpenAlex inherited these records from MAG rather than adding them independently. We found that more than 94% of the articles have a *mag_id*, following a trend similar to the overall article set. This indicates that the vast majority of Chinese journal articles indexed by OpenAlex originate from the period when MAG was in operation. Since the closure of MAG and the launch of OpenAlex in 2021, the inclusion of Chinese journal articles has not increased substantially.

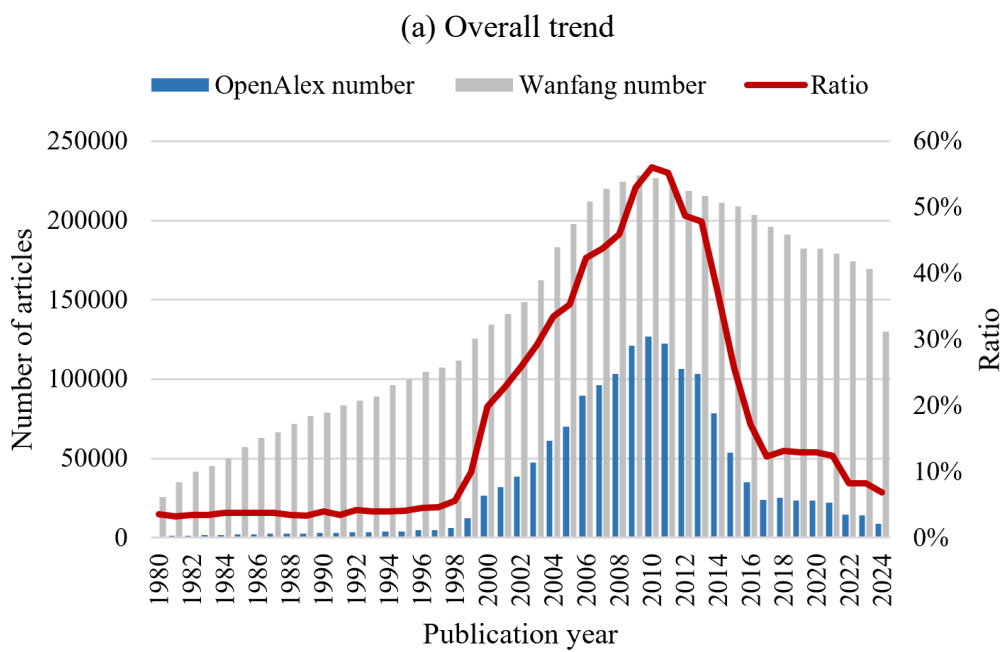

(a) Overall trend

(b) Journal-specific trend

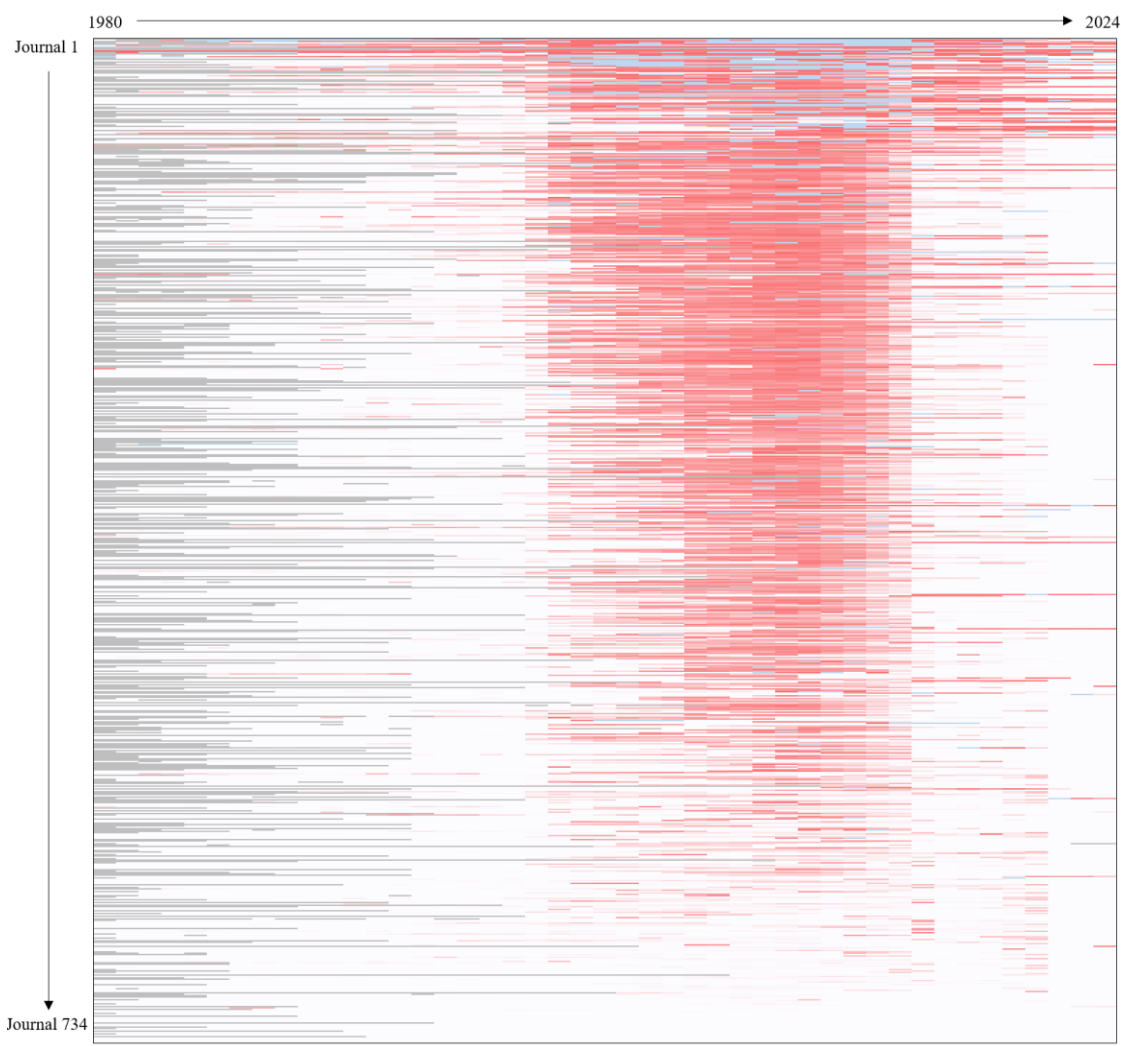

Note: The deeper the red color, the larger the ratio. Cells with ratios greater than 1 are represented in blue. Cells representing units with no articles in the respective year according to Wanfang Data are shaded in dark gray.

Figure 1 Ratio of the articles covered in OpenAlex to Wanfang Data for 734 Chinese journals between 1980 to 2024

However, it is important to note that the ratios in Figure 1 do not precisely reflect true coverage. A thorough manual inspection of journal volumes and our DOI-based analysis revealed several data quality issues in OpenAlex, including missing metadata, incorrect field tagging, and duplicate works. As a result, the articles mapped to journals by OpenAlex may not be perfectly aligned with the corresponding local data.

To be more specific, we observed that some articles indexed in OpenAlex lack journal information, which prevents retrieval via p-ISSN. This observation is consistent with our finding that 192,108 articles with valid DOIs in Wanfang Data could not be retrieved via p-ISSN but were matched through DOI. One such case is the paper titled "Can rock magnetic fabric reveal strain? Case studies of Early Triassic limestones from South China Block"[4] published in the 10th issue of *Chinese Journal of Geophysics* in 2009. In OpenAlex, the source field for this article is empty, which contributes to underestimation in coverage ratios.

Conversely, OpenAlex sometimes conflates journals with similar names, leading to overestimation. For instance, *Acta Geologica Sinica* (ISSN 0001-5717) and *Acta Geologica Sinica - English Edition* (ISSN 1000-9515) are two distinct journals, but OpenAlex incorrectly attributes articles from one to the other. This is demonstrated by the article titled "Active Tectonic Belts in Tibet and Their Control on Earthquakes", which is from the latter but attributed in OpenAlex to the former. This phenomenon is one of the reasons for overestimating the aforementioned ratio, which may cause the existence of blue cells in Figure 1.

A similar issue arises with journals that use the same ISSN but publish different content. For instance, the journals associated with the websites https://journal.xidian.edu.cn/xdxb/CN/1001-2400/home.shtml and https://xadzkjdx.cn/ exist as a single journal entity in OpenAlex, making it hard to distinguish their respective publications. Although these two journals advertise the same ISSN, they in reality are not related and publish different articles. This suggests a potential issue where overseas-registered journals use the ISSN numbers of Chinese domestic journals, raising concerns about misuse or even legitimacy. Specifically, this highlights a limitation in OpenAlex's ability to effectively distinguish between legitimate and fraudulent journals.

In addition, by analyzing the location information of works in OpenAlex[5], we found that, apart from journal websites, data in OpenAlex comes from aggregation platforms such as CNKI in China and J-GLOBAL in Japan. The data quality of these platforms directly influences the quality of the data retrieved by OpenAlex. For example, J-GLOBAL may index the same Chinese journal article in both English and Japanese, resulting in duplicate works in OpenAlex. Such inconsistencies highlight the challenges of verifying and harmonizing web-scraped data for open research information sources.

The examples described above likely reflect broader patterns rather than isolated cases from our sample and they point to systematic issues in OpenAlex with broader implications for the entire database. In the next section, we examine the completeness and accuracy of metadata fields of individual articles to further assess these issues.

## 5. Metadata quality of Chinese journal articles in OpenAlex

To answer the third research question, this section evaluates the completeness and accuracy of selected metadata fields for Chinese journal articles covered in OpenAlex. Completeness refers to the proportion of articles in OpenAlex that contain information for a specific field. This means, whether the field is present and populated. This concept is distinct from accuracy, which refers to the proportion of articles where the content of the field in OpenAlex corresponds to the information available in the local Wanfang Data source.

### 5.1 Metadata completeness

Table 3 presents the completeness of key metadata fields for the Chinese journal articles as found in OpenAlex. All articles in the sample set include publication year information. Since we retrieved the journal articles in this study using ISSNs and excluded articles with missing titles, the completeness for the title and ISSN fields is also 100%. Most articles also include author information. However, bibliographic details such as volume, issue, and page numbers are often missing.

Table 3 Metadata completeness for Chinese journal articles in OpenAlex

| Metadata field | Number of articles in OpenAlex for which the metadata field is populated | Metadata completeness in OpenAlex |
| --- | --- | --- |
| DOI | 418,811 | 27% |
| Title | 1,545,929 | 100% |
| ISSN | 1,545,929 | 100% |
| Year | 1,545,929 | 100% |
| Volume | 554,819 | 36% |
| Issue | 560,639 | 36% |
| First page | 544,270 | 35% |
| Last page | 507,780 | 33% |
| Authors | 1,536,991 | 99% |
| Affiliation strings | 1,046,425 | 68% |
| Institutions | 682,471 | 44% |
| Grants | 624 | 0% |
| References | 104,708 | 7% |

The availability of Digital Object Identifiers (DOIs) for the articles in the sample set is remarkably low[6]. A likely contributing factor is that DOI registration agencies in China do not provide standardized open metadata services comparable to those offered by Crossref (Wang et al., 2018). This limitation makes it much harder for OpenAlex to achieve full completeness for DOI information. The low DOI completeness also means that matching OpenAlex and Wanfang data at the article level cannot rely solely on DOI. To evaluate the accuracy of metadata fields (Section 5.3), more advanced and flexible matching strategies are required (Section 5.2).

Another important issue to note is that OpenAlex provides two types of author affiliation information: raw affiliation strings obtained directly from upstream data sources or article landing pages and PDFs, and lists of standardized institutions linked by OpenAlex based on those raw affiliation strings. As can be observed from Table 3, the proportion of Chinese journal articles in OpenAlex containing affiliation information is relatively low. Only 68% of the articles have raw affiliation strings and only 44% have linked institutions. This highlights limitations in OpenAlex's ability to retrieve, process, and unify affiliation information for Chinese journal articles. In the subsequent analysis, we focus primarily on the linked institutions, as this is the most convenient and directly usable affiliation information for users.

**5.2 Article matching**

To assess the metadata accuracy of Chinese journal articles in OpenAlex compared to Wanfang Data, we used a rule-based approach to match the articles retrieved from OpenAlex to their corresponding records in Wanfang Data. The approach consists of five matching rules, applied in descending order of strictness:

(1) Match by *DOI*: This rule uses DOI, which uniquely identifies scholarly works. Matching by DOI ensures the highest precision and is therefore applied first.

(2) Match by *Title + ISSN + Year + Issue*: Within a specific journal issue, it is uncommon for two articles to share the same title. Therefore combining title, ISSN, year, and issue number provides a reliable match.

(3) Match by *Title + ISSN + Year*: Due to the frequent absence of issue information in many OpenAlex records, this rule omits the issue field while retaining reasonable precision. This rule still ensures high quality matches, as long as general, non-research works such as call-for-paper notices and tables of contents are excluded.

(4) Match by *Title + ISSN*: This matching rule removes the year constraint to account for potential errors in year metadata in OpenAlex. It assumes that journals rarely publish multiple articles with identical titles.

(5) Match by *Title* only: The least strict rule matches articles based solely on their titles. This rule is particularly useful in cases where journals have changed their ISSN over time or where OpenAlex has incorrectly linked an article to the wrong source. However, caution is required when interpreting the results of this matching rule, as it may lead to ambiguous matches, especially for common or generic titles.

It should be noted that, in light of the growing trend of internationalization in academic publishing, some articles published in Chinese journals include bibliographic information in English. Accordingly, we employed both Chinese and English title information for the title-based matching process[7]. During this process, we observed formatting issues in the *Title* field of OpenAlex that could lead to unsuccessful matches. To address this, we applied a series of preprocessing steps to the titles. Specifically, we performed encoding operations (e.g., converting "&" to "&") and employed fuzzy matching techniques by removing punctuation marks (including spaces) from the titles prior to matching.

Using the matching rules introduced above, a total of 1,394,154 articles (90%) were successfully matched. Using the first matching rule, a total of 282,105 articles (18%) were matched based on DOI. This relatively low percentage can be partly attributed to the limited population of the DOI field (27%). Another contributing factor is the presence of errors in DOI metadata. We manually checked 50 randomly extracted articles with a DOI in OpenAlex that failed to match, and found that some records contained incorrect DOIs in OpenAlex[8]. Most of the remaining mismatches were due to DOIs in Wanfang Data being registered through ISTIC, whereas OpenAlex often includes DOIs from other registration agencies. As a result, the DOIs in both sources being incompatible and articles in OpenAlex cannot be matched to the corresponding records in the local Wanfang Data. Using the second matching rule, an additional 196,519 articles (13%) were successfully matched. This low percentage can be explained by the limited availability of issue information in OpenAlex. Using the third matching rule, we successfully matched an additional 859,362 articles (56%), making it the primary method of matching. Using the fourth and fifth matching rules, 9,029 (1%) and 47,139 (3%) articles were matched, respectively.

The remaining 151,775 unmatched articles (10%) can be attributed to several factors. One such factors is that OpenAlex obtains data from a wide range of aggregation platforms hosted in different countries. For instance, according to the location links, there are over 18,000 documents (1%) in our sample from were retrieved from the J-GLOBAL platform in Japan. As a result, some articles in our dataset have titles that are neither in Chinese nor English but in Japanese and lack a DOI (e.g., https://openalex.org/W2742465672), making them difficult to match with Wanfang Data.

### 5.3 Metadata completeness and accuracy for matched articles

Table 4 presents the completeness and accuracy of selected metadata fields for the articles matched using the different rules. The completeness and accuracy of specific fields vary across the matching groups. More specifically:

- The publication year field is complete and accurate, although a few labeling errors exist. It is worth noting that the 0% accuracy for Rule 4 is expected, as articles with correct year information would have been matched by Rule 3. Manual inspection of records with incorrect year information revealed that some discrepancies are minor, often due to delays between online release and formal publication, or between publication and indexing on certain platforms. For instance, the article https://doi.org/10.16511/j.cnki.qhdxxb.2015.24.011 is recorded with a publication year of 2016, while the actual publication occurred in 2015. OpenAlex indexed the paper with the year 2016. In other cases, larger discrepancies were observed, that cannot easily be explained. For instance, the article "Effect of genistein on cell cycle and apoptosis in PC12 cells transfected with APP695MT gene" is published in 2012 according to official website of its journal but indexed with a publication year of 2026 in OpenAlex.

- In contrast, issue numbers and page information is often missing, likely because the primary Chinese data sources used by OpenAlex (e.g. CNKI English version) do not provide these fields. Nevertheless, when present, the accuracy of these fields is generally acceptable. Errors in issue numbers may also stem from irregular publication practices in Chinese domestic journals. For instance, *Chinese Science Bulletin* publishes three issues per month and occasionally combines two issues. Wanfang sometimes assigns a new issue number (e.g., Z1) to the such combined issues, leading to discrepancies with the issue numbers recorded by OpenAlex.

- Due to differences in language and formatting standards for author name indexing between OpenAlex and Wanfang, precise matching of author names of individual articles is challenging. Therefore, this study uses the number of authors as proxy for evaluating accuracy. Although OpenAlex provides author lists for most Chinese journal articles, the accuracy of author counts is relatively low. Under Rule 3, for instance, the accuracy rate is only 15%. Manual examination revealed that many articles matched by Rule 3 originate from the CNKI English version platform, and only the first author is listed in the majority of these articles. This discrepancy may be due to limitations in the parsing algorithms used for automated data extraction. Additionally, even when the number of authors is correct, the sequence in OpenAlex is often incorrect and the first author is often incorrectly listed as the last author.

- For institutional information, we apply a similar approach to that used for author information by focusing on the number of institutions. Completeness of institutional information is lower compared to that of author information. This is consistent with earlier findings on missing institutional metadata in OpenAlex (Zhang et al., 2024). Accuracy of institution counts ranges between 40% and 80%, indicating a relatively low level reliability.

- Finally, the cited reference information in OpenAlex shows low completeness and accuracy for Chinese journal articles. This is consistent with previous evaluations (Gusenbauer, 2024). Only a small proportion of articles contain reference information, and the accuracy of reference counts is limited. This highlights that there are significant data quality issues in OpenAlex regarding references cited in Chinese journal articles.

Table 4 Metadata completeness and accuracy for matched articles in OpenAlex

|  | Rule 1: match by *DOI* (282,105, 18%) | | Rule 2: match by *Title+ISSN+Year +Issue* (196,519, 13%) | | Rule 3: match by *Title+ISSN+Year* (859,362, 56%) | | Rule 4: match by *Title+ISSN* (9,029, 1%) | | Rule 5: match by *Title* (47,139, 3%) | | Total (1,394,154, 90%) | |
|---|---|---|---|---|---|---|---|---|---|---|---|---|
|  | C | A | C | A | C | A | C | A | C | A | C | A |
| Year | 100% | 95% | 100% | 100% | 100% | 100% | 100% | 0% | 100% | 97% | 100% | 98% |
| Issue | 97% | 99% | 100% | 100% | 1% | 76% | 94% | 97% | 17% | 93% | 35% | 99% |
| First page | 94% | 82% | 98% | 93% | 1% | 92% | 85% | 85% | 16% | 90% | 34% | 87% |
| Last page | 91% | 68% | 93% | 76% | 1% | 48% | 80% | 81% | 16% | 88% | 35% | 72% |
| Author count | 99% | 83% | 99% | 89% | 100% | 15% | 99% | 74% | 100% | 40% | 100% | 40% |
| Institution count | 33% | 47% | 22% | 53% | 56% | 62% | 13% | 55% | 52% | 75% | 46% | 59% |
| Reference count | 18% | 5% | 12% | 6% | 0% | 11% | 39% | 8% | 1% | 6% | 6% | 5% |

Note: "C" denotes the completeness, and "A" denotes the accuracy.

Apart from the fields discussed above, we also observed significant unreliability in the language field in OpenAlex. All articles in our sample are from journals that publish in Chinese. However, based on an analysis of the 1,537,378 articles from our sample with language information in OpenAlex, only 5% are labeled as Chinese, while 92% are labeled as English, and a small proportion are labeled as languages such as Korean, Vietnamese, or Danish. The inaccuracy of language metadata in OpenAlex and the underestimation of languages like Russian and Chinese have been confirmed in previous studies (Céspedes et al., 2025) and is also officially acknowledged by OpenAlex (2025). This issue has since been improved (Demes, 2025), but it raises concerns for conducting analyses related to language diversity and similar aspects.

### 5.4 Disciplinary differences in metadata completeness and accuracy for matched articles

We further examined differences in metadata completeness and accuracy across disciplines. As shown in Figure 2, journals in the life and natural sciences exhibit higher completeness for metadata fields such as issue numbers and page ranges. This pattern reflects the greater standardization and internationalization of publishing practices in these fields. In contrast, journals in the humanities and social sciences tend to show higher accuracy in author and institution counts. This may be explained by smaller co-authorship teams in these disciplines (Lewis et al., 2012; Melin, 2000), which reduces the likelihood of errors in author and affiliation data.

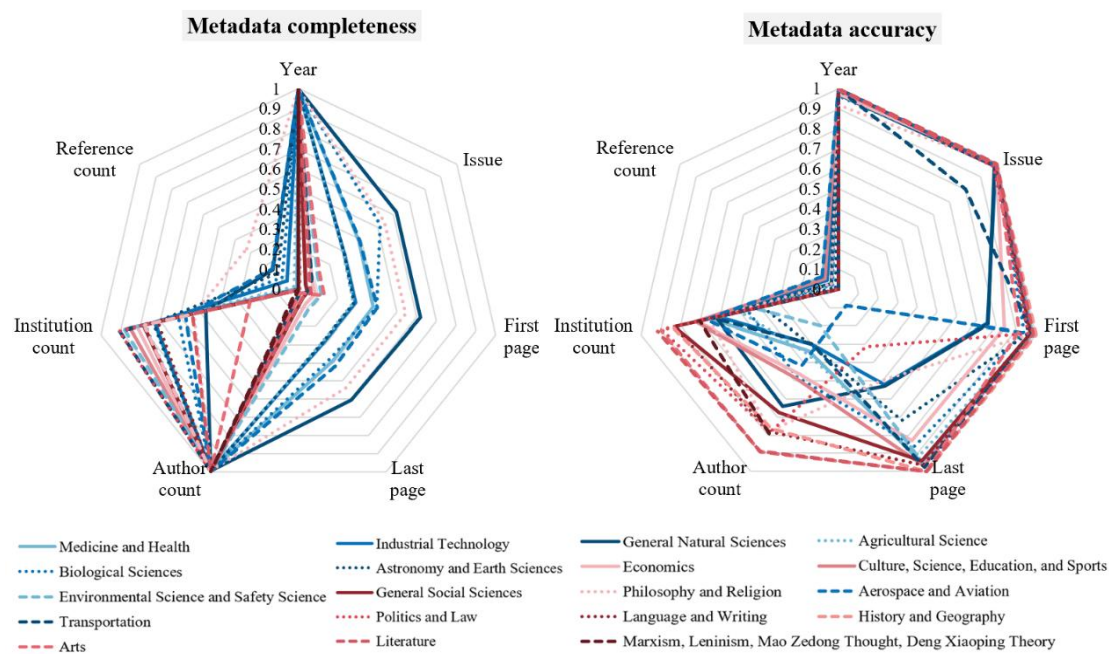

Figure 2 Metadata completeness and accuracy for matched articles in OpenAlex by discipline

## 6. Conclusions and discussion

By comparing the journal and article metadata available in the open data source OpenAlex with those from the local data provider Wanfang Data, this study shows that OpenAlex currently includes just over one-third of Chinese core journals, with an even lower proportion of articles from these journals and notable variations across disciplines and publication years. At the individual article level, our analysis further reveals widespread metadata quality issues such as missing bibliographic information and incomplete author affiliation and citation data. Furthermore, the accuracy of details such as page numbers, authors, institutions, and citations is also low. Although OpenAlex offers broader coverage and inclusivity compared to other proprietary data sources, the results of this study indicate that there is substantial room for improvement in both the coverage of Chinese journal articles and the quality of the associated metadata. These findings highlight current limitations in OpenAlex's ability to provide a fully inclusive view of global scholarly communication and ensure fairness across countries when used for research evaluation and related activities. At the same time, the unique characteristics of China's academic publishing system make achieving these improvements challenging. Nevertheless, these limitations and challenges present opportunities for future improvements, and we therefore propose several recommendations for key actors in the Chinese publishing system and for open research information sources such as OpenAlex.

First, open data sources with a global focus such as OpenAlex should strengthen their strategies for aggregating and integrating data from diverse upstream data sources. Our analysis shows that roughly two-thirds of Chinese journal and article data in OpenAlex comes from local data aggregation platforms such as CNKI, with the remainder sourced from other channels, including Crossref, journal websites, and foreign data aggregation platforms. While this diversity can enhance coverage and inclusivity, it also introduces risks such as inconsistent formats and duplicate records. To improve usability and trust, it would be beneficial for OpenAlex to strengthen its assessment of the reliability of upstream data sources and monitor whether platforms such as CNKI are fully indexed or only partially integrated, especially given recent declines in CNKI-indexed articles. In addition, establishing clear and transparent strategies for data integration is recommended. This includes defining protocols for handling conflicting information provided by upstream data sources and for cases where upstream data sources do not provide certain fields. It would also be helpful for OpenAlex to indicate when a field is empty whether this is because the data could not be retrieved or because the original work does not contain such information. For instance, in the case of author affiliations and cited references, users currently cannot determine whether a work lacks this information or whether OpenAlex was unable to retrieve it. Another recommendation is to make not only the origin of the raw data visible but also provide access to the raw data itself alongside the cleaned and unified version created by OpenAlex. Including provenance information would allow users to trace issues more easily, enabling faster community corrections, and help users better assess data usability for their purposes.

Second, DOI registration practices in China require further development. At present, there are 12 DOI registration agencies worldwide, among which ISTIC&Wanfang and CNKI are the two providing DOI services for Chinese-language journals[9]. Initially, ISTIC&Wanfang provided free registration services. CNKI later became involved and took over some registration responsibilities, but their number of registrations for journal articles and dissertations remains significantly lower than those handled by ISTIC&Wanfang[10]. Each journal can obtain a DOI for an article through one of the two agencies. However, the registration system for DOIs in China is still not fully developed and does not yet fully leverage DOI's role as a unique identifier. On the one hand, the registration process for DOIs in China lacks comprehensive automatic norms. Many Chinese journals do not activate DOIs upon publication. Instead, they adopt a mechanism of pre-assignment followed by registration and activation at a later stage. On the other hand, the metadata required to register a DOI in China is only limited to some essential fields defined by national standards. The metadata fields filled by the DOI registration agencies in China are significantly less rich than those of Crossref[11], and detailed information on keywords, abstracts, and cited references, for instance, is not accommodated. Furthermore, the DOI agencies in China do not offer easy and open access to the data by making snapshots or APIs available so far, like Crossref and DataCite. These limitations currently pose significant challenges for the integration of Chinese local data in international data sources like OpenAlex and require coordinated efforts among multiple stakeholders.

Third, local data aggregators such as Wanfang Data, especially those offering identifier registration services, should act as reliable intermediaries between raw data producers and downstream users and platforms. Better coordination among domestic providers is encouraged to avoid inconsistencies, such as journals registering DOIs with both ISTIC and CNKI simultaneously, which undermines identifier uniqueness and complicates data integration downstream. International standards that accommodate regional and disciplinary differences could help local data aggregators deliver higher-quality, more consistent, and easier-to-integrate data to downstream platforms like OpenAlex.

Finally, journals and publishers, as the original sources of publication metadata, set the upper limit of the data quality. Beyond disseminating high-quality research, they must also maintain the integrity and credibility of academic publishing. However, our analysis reveals certain bad practices such as the misuse of ISSNs and the inclusion of multiple DOIs for the same article, indicating shortcomings in journal management practices that ultimately affect metadata quality. This challenge is more serious in China, where the majority of journals are managed and published by a large number of institutions. According to the statistics, China's 5,211 scientific journals are owned by 3,217 institutions, with 76.87% owning only one journal; and they are published by 4,470 institutions, with 96.04% publishing only one journal (Service Center for Societies of CAST, 2024). Strengthening standardization and international interoperability of Chinese publishing processes is therefore also critical for improving metadata quality and enhancing the visibility of local Chinese journals in the global scholarly ecosystem.

This study focuses exclusively on the journals listed in *A Guide to the Core Journals of China (2023)* and the coverage of these journals and their articles in OpenAlex. Consequently, the findings do not represent the coverage of all types of literature in China and cannot be generalized to other countries or regions. Nevertheless, given the size and authority of the selected journal list and China's significant role in global publishing activities, this study offers valuable insights and an approach that can be expanded and applied to other countries and regions. Furthermore, it is crucial to recognize that data sources such as OpenAlex are continuously evolving. While investigations into metadata quality, including our current study, provide valuable insights at the time of analysis, their relevance may diminish as new data are added and existing records refined. Readers should therefore interpret these findings in the light of the dynamic nature of these data sources and their ongoing development.

## Acknowledgement

The authors would like to acknowledge support from the National Natural Science Foundation of China (Grant Nos. 72374160, L2424104) and the National Laboratory Centre for Library and Information Science at Wuhan University.

## Funding information

National Natural Science Foundation of China, Grant/Award Numbers. 72374160, L2424104.

## Endnotes

[1] http://hxqk.lib.pku.edu.cn/

[2] According to statistics from the official websites, China National Knowledge Infrastructure (CNKI) of TTOD indexes 1,979 of the GCJC journals, and VIP data indexes 1,968 of them.

[3] The OpenAlex data for this study was retrieved via the API in November 2024.

[4] Although our sample consists of Chinese journals and their publications, due to the internationalization of publishing, some journals provide bibliographic information in English.

[5] OpenAlex provides a *landing_page_url* and *pdf_url* field for each paper, which reflects the source from which the paper can be accessed.

[6] Here we report a benchmark figure for comparison: according to the Wanfang records identified through our matching procedure (see Section 5.2 for details), 72% of the articles are associated with a DOI.

[7] Among the articles successfully matched in our study, nearly 90% provided English titles according to Wanfang Data. However, OpenAlex offers titles only in one language based on the information available from its source platform. To account for this, we matched OpenAlex article titles against Wanfang Data using both Chinese and English article titles. If an article was matched successfully in either language, we considered the match valid.

[8] For instance, the DOI of the article corresponding to the link "https://openalex.org/W4205460699" should have the DOI "10.37188/CO.2019-0255".

However, the DOI provided by OpenAlex is "10.37188/co.2020-0255".

[9] https://www-old.doi.org/registration_agencies.html

[10] The total number of registered DOIs for journal articles by ISTIC&Wanfang is 36,329,523 as of April 3, 2025, while CNKI has registered 19,087,061 as of April 10, 2025.

[11] The metadata provided by ISTIC includes journal title, volume, issue, ISSN, publication year, first page, last page, DOI, author, organization, and article title. In contrast, Crossref provided more information including abstract, references, publisher, funder, and more (https://www.crossref.org/documentation/).